\documentstyle[amstex,amssymb,righttag,verbatim,12pt]{article}
\newtheorem{th}{Theorem}

\newtheorem{lem}{Lemma}
\newcommand{\res}{\operatorname{res}}
\begin{document}

\title{\Large\sc The Constrained MKP Hierarchy
and the\\ Generalized Kupershmidt-Wilson Theorem}

\author{Q. P. Liu\thanks{On leave of absence from
Beijing Graduate School, CUMT, Beijing 100083, China}
\thanks{Supported by {\em Beca para estancias temporales
de doctores y tecn\'ologos extranjeros en
Espa\~na: SB95-A01722297(SPAIN) and National Natural Science Foundation(CHINA)}}
  \\
 Departamento de F\'\i sica Te\'orica,\\ Universidad Complutense\\ 
E28040-Madrid, Spain.}

\date{}

\maketitle

\newcommand{\ba}{\begin{array}}
\newcommand{\ea}{\end{array}}
\def\p{\partial}
\def\alf{\alpha}
\def\bi{\beta}
\def\es{\epsilon}
\def\la{\lambda}
\def\fr{\frac}
\def\dta{\delta}
\def\l{\left}
\def\r{\right}
\def\texs{\textstyle}
\def\ds{\displaystyle}
\def\G{{\delta G\over \delta{\cal L}}}
\def\F{{\delta F\over \delta{\cal L}}} 
\def\g{{\delta G\over \delta{L}}}
\def\f{{\delta F\over \delta{L}}} 
\def\H{{\delta H\over \delta{\hat{\cal L}}}} 
%\maketitle
\vspace{0.5in}\begin{center}
\begin{minipage}{5in}
{\bf ABSTRACT}\hspace{.2in} 
The constrained Modified KP hierarchy is considered from the 
viewpoint of modification. It is shown that its
second Poisson bracket, which has a rather complicated form, is
associated to a vastly simpler bracket via Miura-type map. The similar
results are established for a natural reduction of MKP. 
\par 
\end{minipage}
\end{center}
\vspace{.5in}
\vfill\eject
The aim of the paper is to construct the generalized Miura map 
and the modification for the constrained modified KP 
hierarchy(cMKP). We shall show that the 
rather complicated second Hamiltonian structure for cMKP
is transformed to a vastly simpler
structure(essentially ${\p\over\p x}$). 

As a transformation between KdV and MKdV systems, Miura map played
important role in the development of the Soliton theory. Due to its
importance, Miura map has been generalized to 
various integrable systems and we here just cite 
the well-known Kupershmidt-Wilson
theorem\cite{kw} for the Gelfand-Dickey(GD) hierarchy, Drinfeld-Sokolov's
results for the integrable systems related to Kac-Moody-Lie algebras\cite{ds},
Miura maps for the hierarchies
from the energy-dependent Schr$\ddot{o}$dinger operator\cite{apf} 
 and 
matrix GD hierarchy\cite{ian}, etc.. Apart from its mathematical
interest, Miura type map is relevant from the viewpoint
of physics. Indeed, in many cases, Miura maps serve as
free field realizations for the corresponding $W$ algebras.

More recently, so-called the constrained KP hierarchy 
attracts much attention(see \cite{cheng}\cite{ov-s} 
and the references there). It is known that
this hierarchy plays a role in theory of matrix models\cite{bx}.
Its modifications
and Miura transformation were considered initially in\cite{lx} for
the concrete cases and in \cite{cheng} for the general case. 
The same sort 
of problem is considered for the $(n,m)^{th}$
-KdV hierarchy\cite{blx}(see also\cite{dk1}\cite{mas}).

We are interested in the constrained modified KP hierarchy.
It is to be remarked that
the MKP is introduced in\cite{kup} and studied extensively in 
\cite{kup1}\cite{ov1}.
 The Lax operator for cMKP is
\begin{equation}
{\cal L}=\p^n +v_{n-1}\p^{n-1}+\cdots +v_0+\p^{-1}v_{-1},
\end{equation}
and the Lax equation reads
\begin{equation}
{d\over dt_{q}}{\cal L}=[({\cal L}^{q\over n})_{\geq 1}, {\cal L}],
\end{equation}
where $\partial=\partial/\partial x$ and 
for any pseudo-differential operator $A$, we assume
$\ds{A=\sum_{i\geq 0} a_i \p^i}+\sum_{i\leq -1} \p^i a_i
=:\sum_{i\geq 0} A_i +\sum_{i\leq -1} A_i $.

As shown by Oevel and Strampp\cite{ov-s}, the cMKP is
a bi-Hamiltonian system 
\begin{equation}
{d\over dt_{q}}{\cal L}={\cal P}_1 \delta{H}_{q+1}
= {\cal P}_2 \delta{H}_{q},
\end{equation}
with the Poisson tensors or Hamiltonian operators ${\cal P}_i$
are defined by
\begin{equation}
{\cal P}_1{\delta H\over \delta{\cal L}}=
\left[\left( {\delta{H}\over\delta{\cal L}}\right)_{\geq 1}, {\cal L}\right] -
\left(\left[{\delta{H}\over \delta{\cal L}}, {\cal L}\right]\right)_{\geq -1},
\end{equation}

%$$
%{\cal P}_1{\delta H\over \delta{\cal L}}=
%\left[\left( {\delta{H}\over\delta{\cal L}}\right)_{\geq 1}, {\cal L}\right] -
%\left(\left[{\delta{H}\over \delta{\cal L}}, {\cal L}\right]\right)_{\geq -1},
%\eqno(\mbox{4})
%$$
\begin{equation}
\begin{aligned}
{\cal P}_2{\delta H\over \delta{\cal L}}=& 
\l({\cal L}{\delta H\over \delta{\cal L}}\r)_{\geq 0}{\cal L}-
{\cal L}\l({\delta H\over \delta{\cal L}}{\cal L}\r)_{\geq 0}-
\l[\l({\cal L}{\dta H\over \dta{\cal L}}\r)_0, {\cal L}\r]+ \\
&+\p^{-1} {\res}\l(\l[{\cal L}, {\dta H\over \dta{\cal L}}\r]\r){\cal L}+
\l[{\cal L}, D^{-1}\l({\res}\l[{\cal L}, 
{\dta H\over\dta{\cal L}}\r]\r)\r], 
\end{aligned}
\end{equation}
%$$
%\displaylines{
%\hskip2.2cm {\cal P}_2{\delta H\over \delta{\cal L}}= 
%\l({\cal L}{\delta H\over \delta{\cal L}}\r)_{\geq 0}{\cal L}-
%{\cal L}\l({\delta H\over \delta{\cal L}}{\cal L}\r)_{\geq 0}-
%\l[\l({\cal L}{\dta H\over \dta{\cal L}}\r)_0, {\cal L}\r]+ \cr
%\hskip5.5cm
%+\p^{-1} {\texs{res}}\l(\l[{\cal L}, {\dta H\over \dta{\cal L}}\r]\r){\cal L}+
%\l[{\cal L}, D^{-1}\l({\texs{res}}\l[{\cal L}, 
%{\dta H\over\dta{\cal L}}\r]\r)\r], \hfill{(5)} \cr }
%$$
\[
H_q={n\over q}\int{{\res}\,{\cal L}^{q\over n} dx},
\]
where $\res$ stands for the usual residue: 
for any operator $A=\sum a_i\p^i$, 
${\res} A=a_{-1}$; also the notion $A_0$ means taking the projection
to $a_0$; the gradient ${\delta H\over \dta {\cal L}}$ is defined as
${\dta H\over\dta{\cal L}} =\sum_{i=-1}^{n-1}\partial^{-i-1}{\dta H\over
\dta v_i}$.

While the first bracket induced by ${\cal P}_1$ is the Lie-Poisson bracket
associated to the relevant algebra, the second one has a rather 
complicated form. Our aim is to get a better understanding of its
structure. We will see that this complicated Poisson tensor 
is Miura-related to a very simple one.
Indeed,
the initial steps along this direction were taken 
in a previous paper\cite{liu}
and there we made a conjecture which now is formulated as the following:  
\begin{th}
Define a Miura transformation between 
the coordinates $(v_{n-1},\dots\\ 
,v_0, v_{-1})$
and $(w_n,\dots, w_0)$ by the following factorization of the Lax
operator ${\cal L} $
\begin{equation}
{\cal L}=\p^{-1}(\p-w_n)(\p-w_{n-1})\cdots(\p-w_0),
\end{equation}
then, the second bracket induced by ${\cal P}_2$ 
\begin{equation}
\{F,G\}=\int{{\res}\left( {\dta F\over \dta{\cal L}}
({\cal P}_2{\dta G\over\dta{\cal L}})\right) }dx,
\end{equation}
is transformed to
\begin{equation}
\{F,G\}=-\int{ \sum^{n}_{i=0}{\dta F\over\dta{w_i}}
({\dta G\over\dta{w_i}})^{\prime}} dx+
\int{\sum^{n}_{i=0}{\dta F\over\dta{w_i}}
\sum^{n}_{j=0}({\dta G\over\dta{w_j}})^{\prime}} dx.
\end{equation}
where $\prime=\partial$ for short.
\end{th}
To make the things easier, we first convert the bracket(7) to more
familiar Gelfand-Dickey brackets as in\cite{huang}. This is the context 
of the following Lemma.
\begin{lem}
Let us define a new operator by
\begin{equation}
L\equiv \p {\cal L}= \p^{n+1}+u_n\p^n+\cdots+u_0,
\end{equation}
then, the following relation between brackets is satisfied
\begin{equation}
\{F,G\}=\{F,G\}_2+\{F,G\}_3,
\end{equation}
where
\begin{equation}
\{F,G\}_2=\int{{\res}{\dta F\over\dta L}\left(
(L{\dta G\over\dta L})_{\geq 0}L
-L({\dta G\over\dta L}L)_{\geq 0}\right) }dx,
\end{equation}
and
\begin{equation}
\{F,G\}_3=\int{{\res} {\dta F\over\dta L}\l[L,D^{-1}
{\res}\l[L,{\dta G\over\dta L}\r]\r] }dx,
\end{equation}
are the so-called second and third Gelfand-Dickey brackets 
associated to
$L$  respectively. $D^{-1}$  denotes an integration: 
$D^{-1}=\int^x dz$; $\{F,G\}$ is given by (7).
\end{lem}
{\em Proof}. Since $ {\dta F}=\int{{\res}({\dta F\over\dta L}{\dta L}})dx=
\int{{\res}({\dta F\over\dta L}\p\dta{\cal L})}dx=
\int{{\res}(\F\dta{\cal L})}dx$, one has
\begin{equation}
{\dta F\over\dta {\cal L}}={\dta F\over\dta L}\p,
\end{equation}
Then,
\[
\l(L{\dta F\over\dta L}\r)_{\geq 0} = 
\l(\p{\cal L}{\dta F\over\dta {\cal L}}\p^{-1}\r)_{\geq 0}
 =
\l({\cal L}{\dta F\over\dta {\cal L}}\r)_{\geq 0}+
\l({\cal L}{\dta F\over\dta {\cal L}}\r)^{\prime}_{\geq 1}\p^{-1}, 
\]
\begin{equation}
\begin{aligned}
\l(L{\dta F\over\dta L}\r)_{\geq 0}L &=
\l({\cal L}{\dta F\over\dta {\cal L}}\r)_{\geq 0}\p {\cal L}+
\l({\cal L}{\dta F\over\dta {\cal L}}\r)^{\prime}_{\geq 1}{\cal L} \\
&=\p\l({\cal L}{\dta F\over\dta {\cal L}}\r)_{\geq 0}{\cal L}-
\l({\cal L}{\dta F\over\dta {\cal L}}\r)^{\prime}_{0}{\cal L}.
\end{aligned}
\end{equation}
\begin{equation} 
L({\dta F\over\dta L}L)_{\geq 0}=
\p{\cal L}({\dta F\over\dta {\cal L}}{\cal L})_{\geq 0}.
\end{equation}
\begin{equation}
\begin{aligned}
{\res}\l[L, {\dta F\over\dta L}\r]&=
    {\res}\left(\p{\cal L}\F\p^{-1}-\F {\cal L}\right)  \\
&={\res}\left(\l[{\cal L},\F\r]-{\cal L}\F
+\p{\cal L}\F\p^{-1}\right) \\
&=
{\res}\l[{\cal L}, \F\r]+\left({\cal L}\F\right)^{\prime}_{0}.
\end{aligned}
\end{equation}

\noindent
Thus, using the above equations(14-16), we obtain
\begin{align*}
\{F,G\}_2+&\{F,G\}_3  =\\
=& {\displaystyle\int} {\res} \l(\F\p^{-1}
                      \l(\p\l({\cal L}\G\r)_{\geq 0}{\cal L}
                       -\l({\cal L}\G\r)^{\prime}_{0}{\cal L}\r.\r.\\[1.6mm]
         &     \l.\l.- \p{\cal L}\l(\G {\cal L}\r)_{\geq 0} 
                       +\l[\p{\cal L}, D^{-1}{\res}\l[{\cal L}, \G\r]\r]+
                          \l[\p{\cal L}, \l({\cal L}\G\r)_0\r]\r)\r)dx \\[1.6mm]
           =&  {\displaystyle\int} {\res}\l(\F {\cal L}\p^{-1}
                       \l(\p\l({\cal L}\G\r)_{\geq 0}{\cal L}-
                          \p{\cal L}\l(\G {\cal L}\r)_{\geq 0}-
                             \p\l[\l({\cal L}\G\r)_0, {\cal L}\r]\r.\r. 
\\[1.6mm]
          &    \l.\l. +\p\l[{\cal L}, D^{-1}{\res} \l[{\cal L}, \G\r]\r]+
                          {\res}\l(\l[{\cal L},\G\r]\r){\cal L}\r)\r) dx 
\\[1.4mm]
         =&   \{F,G\}.  
\end{align*}

\noindent 
This completes the proof the lemma 1. $\Box$

With lemma 1 in hand, we may use the standard 
Kupershmidit-Wilson\cite{kw}\cite{dic}
theorem for the second GD bracket. It states that
\begin{equation} 
\{F,G\}_2=-\int\sum^{n}_{i=0}\l({\dta F\over\dta{w_i}}\r)
\l({\dta G\over\dta{w_i}}\r)^{\prime}dx,
\end{equation}
thus, to prove the theorem 1, we need only to prove the following lemma
\begin{lem}
\[
\{F,G\}_3=
\int{\sum^{n}_{i=0}{\dta F\over\dta{w_i}}
\sum^{n}_{j=0}\l({\dta G\over\dta{w_j}}\r)^{\prime}}dx.
\]
\end{lem}
{\em Proof}.  Using $\dta F=\int{{\res}{\dta F\over\dta{L}}\dta{L}}dx=
\int{\sum^{n}_{i=0}{\dta F\over\dta{w_i}}\dta{w}_i}dx$, we obtain
\begin{equation}
{\dta F\over\dta{w_i}} =-{\res}\l(\p_{i-1}\cdots\p_0{\dta 
F\over\dta{L}}\p_n\cdots\p_{i+1}\r),
\end{equation}
here we used the notion $\p_i=\p-w_i$ for short.

Now,
\noindent
\[
\int\sum^{n}_{i=0}{\dta F\over\dta{w_i}}
\sum^{n}_{j=0}\l({\dta G\over\dta{w_j}}\r)^{\prime}dx= 
\]
\[ 
 =\ds\int {\res}\left(\sum (\p_{i-1}\cdots\p_0\f \p_n\cdots\p_{i+1})            
   \l({\res}\sum(\p_{j-1}\cdots\p_0\g\p_n\cdots\p_{j+1})^{\prime}\r) \right)dx 
\]
\[
=\ds\int {\res}\left(\sum (\p_{i-1}\cdots\p_0\f\p_n\cdots\p_{i+1})
      \l[\p, {\res}\sum (\p_{j-1}\cdots\p_0\g\p_n\cdots\p_{j+1})\r]\r)dx 
\]
\[
=\ds\int {\res} \l(\sum (\p_{i-1}\cdots\p_0\f\p_n\cdots\p_{i+1}) 
\l(\p \l({res} \sum (\p_{j-1}\cdots\p_0\g\p_n\cdots\p_{j+1})\r)\r.\r.   
\]
\[
\ds\l.\l. -\l({\res} \sum (\p_{j-1}\cdots\p_0\g\p_n\cdots\p_{j+1})\r)\p\r)\r) 
dx. 
\]

We replace $\p$ by $\p_i$ in the last expression, which changes nothing,
then we find
$$
\displaylines{
\int\sum^{n}_{i=0}{\dta F\over\dta{w_i}}
\sum^{n}_{j=0}({\dta G\over\dta{w_j}})^{\prime}dx=
\int {\res}\l(\l[{\dta F\over\dta{L}},L\r] 
{\res}\sum (\p_{i-1}\cdots\p_0{\dta G\over\dta{L}}\p_n\cdots\p_{i+1})\r)
,\hfill{(19)} \cr }
$$
\setcounter{equation}{19}
however,
\begin{eqnarray}
\lefteqn{\l({\res}\sum{(\p_{i-1}\cdots\p_0\g\p_n\cdots\p_{i+1}})\r)^{\prime}=} 
\nonumber \\
& & ={\res}\l[\p, \sum{(\p_{i-1}\cdots\p_0\g\p_n\cdots\p_{i+1}}\p)\r]\nonumber 
\\
& & ={\res}
\l[L,\g\r], 
\end{eqnarray}
which implies
\begin{equation}
{\res}\sum{(\p_{i-1}\cdots\p_0{\dta G\over\dta{L}}\p_n\cdots\p_{i+1}})=
D^{-1}{\res}\l[L,{\dta G\over\dta{L}}\r],
\end{equation}
now substituting the above expression into the equation(19), we complete
the proof.$\Box$

Thus, the combination of the last two Lemmas provides us a completed
proof for our theorem 1.

Next let us turn our attention to a special reduction of the general
case. That is, we consider the following Lax operator
\begin{equation} 
{\hat{\cal L}}=\p^n+v_{n-1}\p^{n-1}+\cdots+v_1\p ,
\end{equation} 
It is easy to see
that this is a consistent reduction. The Hamiltonian
structures for the associated hierarchy 
are inherited from the one(4) by Dirac reduction\cite{ov}.
Since we are only
interested in the second one, let us write it down here
\begin{equation}
\begin{aligned}        
{\hat{\cal P}}(\H )  =&
\l({\hat{\cal L}}\H\r)_{\geq 0}{\hat{\cal L}}-
{\hat{\cal L}}\l(\H {\hat{\cal L}}\r)_{\geq 1}-
\l(\H {\hat{\cal L}}\r)_0 {\hat{\cal L}} \\
& -{\hat{\cal L}}\p^{-1}{\res}\left(\l[\H,{\hat{\cal L}}\r]\right)-
\l[D^{-1}\left({\res}\l[\H,{\hat{\cal L}}\r]\r),{\hat{\cal L}}\r],
\end{aligned}
\end{equation}
where ${\dta H\over \dta {\hat{\cal L}}}=\sum_{i=1}^{n-1}
\partial^{-(i+1)}{\dta H\over \dta v_i}$.

We define a bracket with ${\hat{\cal P}}$ as
\begin{equation}
\{F,G\}=\int {\res}\left({\dta F\over\dta{\hat{\cal L}}}{\hat{\cal P}} 
{\dta G\over\dta{\hat{\cal L}}}
\right)dx.
\end{equation}

This bracket is also complicated and its structure is revealed
by the following lemma:
\begin{lem}
\begin{equation} 
\{F,G\}=\{F,G\}_2-\{F,G\}_3,
\end{equation}
where the left side of the equation is the second and third GD 
brackets associated with the operator ${\hat L}={\hat{\cal L}}\p^{-1}$.
\end{lem}
{\em Proof}. Using
the identity $ \p \H={\dta H\over\dta{\hat{L}}}$ and the rest is similar
to the proof of  lemma 1. $\Box$

At this point, we have the following theorem
\newpage
\begin{th}. 
Let the Miura map be given by the following factorization of 
the Lax operator ${\hat L}$ 
\begin{equation}
{\hat L}=(\p-{\hat{w}}_{n-1})\cdots(\p-{\hat{w_{1}}}).
\end{equation}
Then in the coordinates $\{{\hat{w}}_i\}$,
 the Poisson tensor ${\hat{\cal P}}$  defined by (23)
 is the following simple $(n-1)\times (n-1)$ matrix operator:
\begin{equation}
-\left(\ba{cccc}2&1&\cdots&1\\ 1 &\ddots&\ddots&\vdots\\
\vdots&\ddots&\ddots&1\\ 1&\cdots&1&2\ea\right)\p.
\end{equation}
\end{th}

\noindent
{\em Remark}: We know that a Poisson tensor
will define a Poisson algebra. So, our theorem 1 and theorem 2
provide free field realizations for the corresponding
Poisson algebras. This may be interesting for the theory of 
$W$ algebra.

To conclude the paper, we give here one example.

\noindent
{\em Example}. $n=3$ case: ${\hat{\cal L}}=\p^3+v_2\p^2+v_1\p$. 
The Lax equation
\[
{\hat{\cal L}}_t=[({\hat{\cal L}}^{2\over3})_{\geq 1}, {\hat{\cal L}}],
\]
gives
\begin{align*}
v_{2_{t}}&={1\over3}(-3v_{2_{xx}}+6v_{1_{x}}-2v_2v_{2_{x}}),\\
v_{1_{t}}&={1\over3}(-2v_{2_{xxx}}+3v_{1_{xx}}+2v_2v_{1_{x}}-2v_1v_{2_{x}}
-2v_2v_{2_{xx}}).
\end{align*}

The second Poisson tensor is
\[
{\hat{{\cal P}}}=\left(\ba{cc}-6\p&3\p(\p-v_2)\\
          -3(\p+v_2)\p &2\p^3+v_1\p+\p v_1+2v_2\p^2-2\p^2v_2-2v_2\p v_2
          \ea\right).
\]

According to the  theorem 2, the modified Poisson tensor 
and the transformation between the fields are respectively
\[
-\left(\ba{cc}2\p&\p\\
              \p&2\p\ea\right)
,\qquad\qquad v_2=-(w_2+w_1), \quad v_1=w_1w_2-w_{1_{x}}
\]
this of course can be verified by a simple calculation.
\bigskip

{\em Acknowledgment}.
I would like to thank Boris Kupershmidt and Ian Marshall
for several comments. 
\par

\end{document}